\documentclass[prl,aps,showpacs,reprint,superscriptaddress,amsmath,amssymb,floatfix,10pt]{revtex4-1}
\usepackage{amsmath,amssymb,graphicx}
\usepackage[colorlinks=true, citecolor=blue, linkcolor=blue, urlcolor=blue]{hyperref}
\usepackage{array}

\begin{document}

\title{Incoherent Diffractive Imaging via Intensity Correlations of hard X-rays}

\author{Anton Classen} 
\affiliation{Institut f\"ur Optik, Information und Photonik, Universit\"at Erlangen-N\"urnberg,
	91058 Erlangen, Germany}
\affiliation{Erlangen Graduate School in Advanced Optical Technologies (SAOT), Universit\"at Erlangen-N\"urnberg,
	91052 Erlangen, Germany}

\author{Kartik Ayyer} 
\affiliation{Center for Free-Electron Laser Science, Deutsches Elektronen-Synchrotron DESY, Notkestr.~85, 22607 Hamburg, Germany}

\author{Henry Chapman} 
\affiliation{Center for Free-Electron Laser Science, Deutsches Elektronen-Synchrotron DESY, Notkestr.~85, 22607 Hamburg, Germany}
\affiliation{Department Physik, Universit\"at Hamburg, Luruper Chaussee 149, 22761 Hamburg, Germany}
\affiliation{The Hamburg Centre for Ultrafast Imaging, Luruper Chaussee 149, 22761 Hamburg, Germany}

\author{Ralf R\"ohlsberger} 
\affiliation{The Hamburg Centre for Ultrafast Imaging, Luruper Chaussee 149, 22761 Hamburg, Germany}
\affiliation{Deutsches Elektronen-Synchrotron DESY, Notkestr.~85, 22607 Hamburg, Germany}

\author{Joachim von Zanthier} 
\affiliation{Institut f\"ur Optik, Information und Photonik, Universit\"at Erlangen-N\"urnberg,
	91058 Erlangen, Germany}
\affiliation{Erlangen Graduate School in Advanced Optical Technologies (SAOT), Universit\"at Erlangen-N\"urnberg,
	91052 Erlangen, Germany}

\begin{abstract}
Established x-ray diffraction methods allow for high-resolution structure determination of crystals, crystallized protein structures or even single molecules. While these techniques rely on coherent scattering, incoherent processes like Compton scattering or fluorescence emission---often the predominant scattering mechanisms---are generally considered detrimental for imaging applications. Here we show that intensity correlations of incoherently scattered x-ray radiation can be used to image the full 3D structure of the scattering atoms with significantly higher resolution compared to conventional coherent diffraction imaging and crystallography, including additional three-dimensional information in Fourier space for a single sample orientation. We present a number of properties of incoherent diffractive imaging that are conceptually superior to those of coherent methods.
\end{abstract}

\maketitle

The advent of accelerator-driven x-ray free-electron lasers (FEL) has opened new avenues for high-resolution x-ray structure determination via coherent diffractive imaging (CDI) methods that go far beyond conventional x-ray crystallography \cite{Chapman2006a,Chapman2011a,Seibert2011,Loh2012,Kupitz2014,Takahashi2013,Barke2015,Neutze2000,Aquila2015,Barty2013,Ayyer2016}. In these methods it is assumed that a fixed phase relation between the incoming and scattered photons exists and the first-order coherence of the radiation field is maintained throughout the imaging procedure. This produces a stationary interference pattern upon measurement of large numbers of photons, a central paradigm of the field since its foundation more than hundred years ago. Incoherence induced by, e.g., time-varying wavefront distortions or incoherent scattering processes like fluorescence emission or Compton scattering, is generally considered detrimental in this approach, as the scattered photons on average generate a constant intensity distribution producing a background that reduces the fidelity of CDI measurements \cite{Slowik2014,Gorobtsov2015,Chapman2010}. 

The situation is fundamentally altered if the photons are recorded within their coherence time $\tau_c$, i.e., a time interval short with respect to the temporal phase fluctuations of the radiation field. Over such short times the relative phases of the scattered photons can be considered as stable allowing to observe a stationary fringe pattern. The pattern will fluctuate and spatially vary over times longer than $\tau_c$, yet the autocorrelation of the intensity distribution calculated for each short exposure is insensitive to the spatial pattern variations and will continuously build up when averaging over many short measurements.

It was this approach that led Hanbury Brown and Twiss (HBT) to their landmark experiment in stellar interferometry to overcome atmospheric fluctuations and determine the diameter of stars via intensity correlations \cite{HBT1956-2}. Based on the discovery of photon bunching of thermal light \cite{HBT1956-1} the HBT experiment initiated a paradigm shift towards a quantum statistical description of light and is nowadays regarded as one of the founding pillars for the development of modern quantum optics \cite{Glauber2006}. The generalized concept of optical coherence \cite{Glauber1963-2} has become a wide-spread and powerful tool in various fields of physics, ranging from stellar interferometry to nuclear collisions \cite{Baym1998,Padula2005}. Recently, even intensity correlations of order $m > 2$ have been measured, allowing for imaging with sub-Abbe resolution \cite{Thiel2007,Dertinger2009,Oppel2012a,Schwartz2013,Genovese2014a,Classen2016,Schneider2017}. 

In this Letter we propose that intensity correlations of incoherently scattered photons can be used to determine the 3D arrangement of atoms in crystals and molecules. For example, in the case of $K$-shell fluorescence photons from transition metal atoms the coherence time is given by their radiative lifetime (for Fe atoms $\tau_c = h /\Gamma = 2.6$\,fs for a linewidth of $\Gamma = 1.6$\,eV \cite{Krause1979}). Excitation of the atoms with femtosecond pulses from current x-ray FELs and measurement of the scattered radiation shot-by-shot fulfills the condition of a fixed phase relation for each exposure and can be applied to derive the 3D structure of the fluorescing atoms with atomic resolution. This approach, which we call incoherent diffraction imaging (IDI), opens fundamentally new strategies for x-ray structure determination based on the measurement of incoherent radiation. 

\begin{figure}[t]%
\centering
\includegraphics[width=1.0 \linewidth]{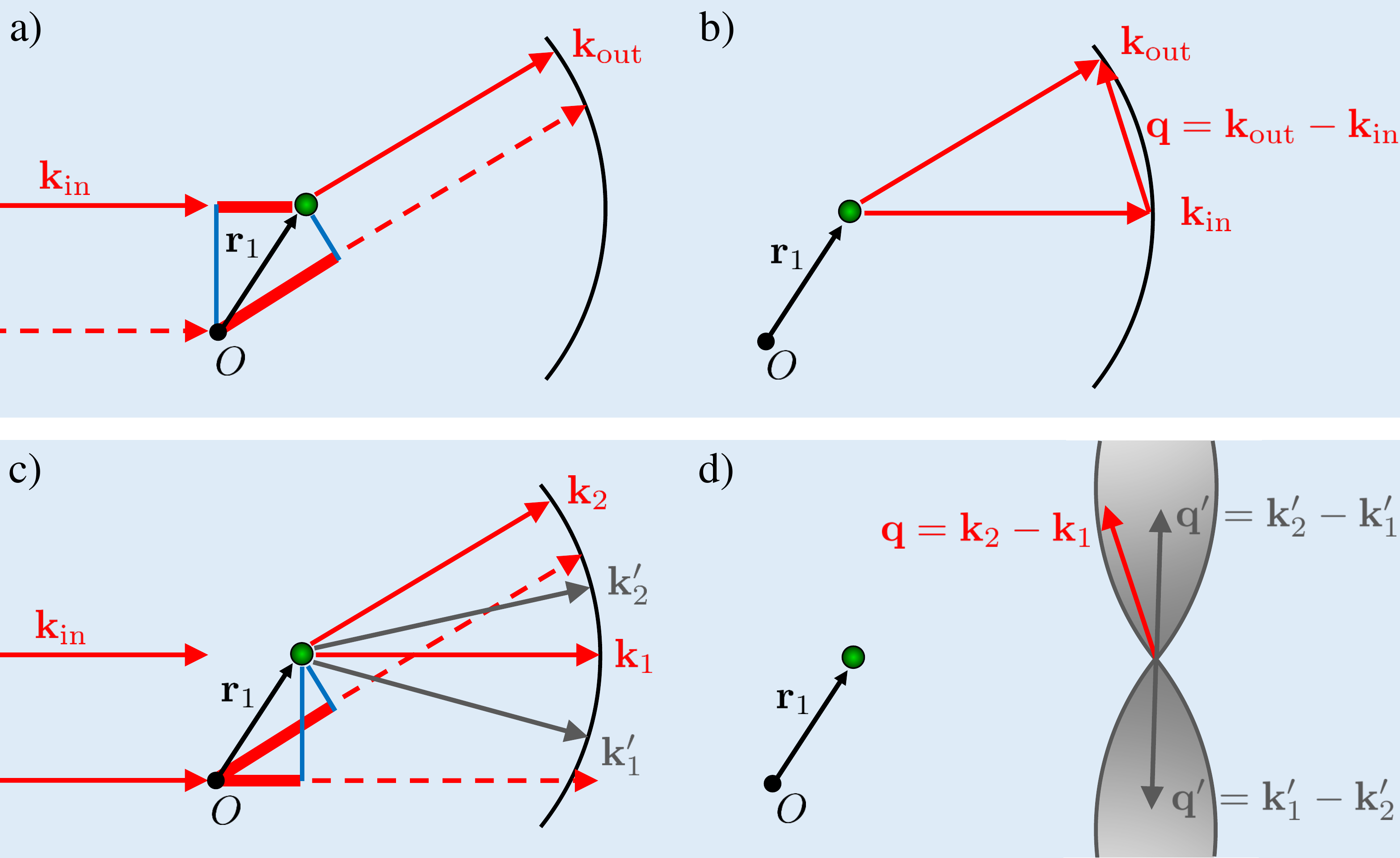}
\caption{(color online) a) Phase acquired by a photon with wave vector $\mathbf{k}_\text{in}$ upon coherent scattering by an atom at  $\mathbf{r}_i$ into direction $\mathbf{k}_\text{out}$, relative to scattering by an atom at the origin, b) Corresponding Ewald sphere construction using the photon momentum transfer $\mathbf{q} = \mathbf{k}_\text{out} - \mathbf{k}_\text{in}$. The black arcs in a) and b) represent the angular extent of a detector in the far field and the 2D Ewald sphere coverage in 3D Fourier space, respectively. c) and d) Phase difference between two photons incoherently scattered by an atom at  $\mathbf{r}_i$, one into direction $\mathbf{k}_2$ and the other one into direction $\mathbf{k}_1$, relative to an atom at the origin. The incoming wave does not transfer any photon momentum on the outgoing wave. However, intensity correlations between outgoing wave vectors $(\mathbf{k}_\text{1}, \mathbf{k}_\text{2})$, $(\mathbf{k}_\text{1}', \mathbf{k}_\text{2}')$, etc.~induce momentum transfers $\mathbf{q} = \mathbf{k}_2 - \mathbf{k}_1$, $\mathbf{q}' = \mathbf{k}_2' - \mathbf{k}_1'$, etc.~which contain information about the object. As a consequence a volumetric 3D coverage of $\mathbf{q}$-values in Fourier space builds up, that reaches out twice as far as the Ewald sphere for common detector geometries.}
\label{fig:scattering}%
\end{figure}

To start with, we assume without loss of generality that the sample under study is composed of an arrangement of $N$ identical atoms with a spatial distribution $S(\mathbf{r}) = \sum_{i = 1}^N \delta(\mathbf{r} - \mathbf{r}_i)$. In CDI the sample is illuminated with photons of wavevector $\mathbf{k}_\text{in}$, and the diffraction pattern is recorded in the far field, yielding the intensity in the direction  $\mathbf{k}_\text{out}$
\begin{equation}
I(\mathbf{\mathbf{k}_\text{out}}) = I_0 P \, |f(\mathbf{q})|^2\, |\tilde{S}(\mathbf{q})|^2 \, .
\label{eq:7}
\end{equation}
Here, $I_0$ is the intensity of the incoming FEL beam, $P$ the polarization factor, $\mathbf{q} = \mathbf{k}_\text{out} - \mathbf{k}_\text{in}$ the photon momentum transfer (see Fig.~\ref{fig:scattering}a), $f(\mathbf{q})$ the atomic form factor that accounts for the electronic charge distribution of a single atom, and $\tilde{S}(\mathbf{q})$ the 3D Fourier transform of $S(\mathbf{r})$, i.e.,
\begin{equation}
\tilde{S}(\mathbf{q}) = FT\{ S(\mathbf{r}) \} = \sum_{i = 1}^N e^{i \mathbf{q} \cdot \mathbf{r}_i}.
\label{eq:5}
\end{equation}
In Eq.~(\ref{eq:5}), $\mathbf{q} \cdot \mathbf{r}_i$ is the phase acquired by a photon with wave vector $\mathbf{k}_\text{in}$ upon coherent scattering by an atom at location $\mathbf{r}_i$ into direction $\mathbf{k}_\text{out}$ relative to scattering by an atom at the origin (see Fig.~\ref{fig:scattering}a). Summation over all scatterers in the object leads to the scattering amplitude $\tilde{S}(\mathbf{q})$, i.e., the strength of coherent diffraction at the particular spatial frequency of the object with wavenumber $|\mathbf{q}|$ and direction $\mathbf{q}$. Note that, due to energy conservation in elastic scattering the vector $\mathbf{q}$ lies on the surface of the so-called Ewald sphere, a 2D shell in 3D Fourier space (see Fig.~\ref{fig:scattering}b). Therefore, the data recorded in a single exposure only displays spatial frequencies of the object that lie on the Ewald sphere. To obtain the object's entire 3D structure additional diffraction patterns with varying orientations $\mathbf{k}_\text{in}$ of the incident beam relative to the object (or vice versa) have to be recorded.

The reconstruction of a 3D image of the object requires the inversion of Eq.\,(\ref{eq:5})
\begin{equation}
S(\mathbf{r}) = FT^{-1} \{ \tilde{S}(\mathbf{q}) \} = \int {\tilde{S}(\mathbf{q}) e^{-i \mathbf{q} \cdot \mathbf{r}} } \, \text{d} \mathbf{q} \, .
\label{eq:9}
\end{equation}
While the modulus of the Fourier amplitudes $|\tilde{S}(\mathbf{q})|$ can be obtained from $\sqrt{I(\mathbf{k}_\text{out})}$, the diffraction phases are missing, resulting in the well-known phase problem. To recover the phases, iterative phase-retrieval algorithms have been developed which successfully resolve this problem \cite{Fienup1982,Elser2003}.

Let us now consider incoherently scattered photons originating from the same set of point-like emitters $S(\mathbf{r})$. For the mathematical derivation of the intensity correlations we use a quantum mechanical treatment. In the far field the positive frequency part of the operator of the outgoing electric field propagating in the direction $\mathbf{k}$ reads
\begin{equation}
\begin{aligned}
\left[ \hat{E}^{(-)}(\mathbf{k}) \right]^\dagger =
 \hat{E}^{(+)} (\mathbf{k}) 
 \sim \sum_{i=1}^N  \ e^{ i \mathbf{k}  \cdot \mathbf{r}_{i}} e^{i \phi_i} \, \hat{a}_i \, ,
\end{aligned}
\label{eq:14}
\end{equation}
where the incoherence of the emission process is incorporated by the randomly and independently varying phases $\phi_i \in [0,2\pi)$, $\hat{a}_i$ denotes the annihilation operator for a photon from emitter $i$, and the geometrical phase $\mathbf{k}  \cdot \mathbf{r}_{i}$ is expressed relative to a photon emitted from the origin (see Fig.\,\ref{fig:scattering}c).  Due to the independently fluctuating phases we obtain for the expectation values $\langle e^{ \pm i \phi_i} e^{ \pm i \phi_j}  \rangle$ = 0 with $i \neq j$. Calculating the intensity scattered into the direction $\mathbf{k}$, i.e., the first-order intensity correlation function $G^{(1)}(\mathbf{k},\mathbf{k}) = \langle \hat{E}^{(-)} (\mathbf{k}) \hat{E}^{(+)} (\mathbf{k}) \rangle$, we obtain \cite{Glauber1963-2}
\begin{equation}
I(\mathbf{k}) = G^{(1)}(\mathbf{k},\mathbf{k}) \sim \sum_{i=1}^N \langle  \hat{a}_i^\dagger  \hat{a}_i \rangle = \sum_{i=1}^N  \langle \hat{n}_i \rangle  \equiv I_\text{tot} \, ,
\label{eq:15}
\end{equation}
where $\langle  \hat{a}_i^\dagger  \hat{a}_i \rangle = \langle \hat{n}_i \rangle$ is the average mode occupation number of emitter $i$ per time interval. In the case of classical light sources, e.g., thermal light sources (TLS), this value can take arbitrary values ranging from $\langle \hat{n}_i \rangle \ll 1$  to $\langle \hat{n}_i \rangle \gg 1$. In the case of single-photon emitters (SPE) we have $\langle \hat{n}_i \rangle \leq 1$ for continuous as well as pulsed excitation. As can be seen from Eq.~(\ref{eq:15}), $I(\mathbf{k})$ is independent of $\mathbf{k}$, therefore for incoherent scattering no information about the spatial source distribution can be gained from $G^{(1)}(\mathbf{k},\mathbf{k})$.

In contrast to $G^{(1)}(\mathbf{k},\mathbf{k})$, the complex degree of coherence $g^{(1)}(\mathbf{k}_1,\mathbf{k}_2)$, i.e., the normalized cross-correlation between two incoherently scattered outgoing electric fields propagating into the directions $\mathbf{k}_1 \neq \mathbf{k}_2$ contains structural information of the source arrangement (c.f. Fig.\,\ref{fig:scattering}c,d)
\begin{equation}
\begin{aligned}
& g^{(1)}(\mathbf{k}_1,\mathbf{k}_2)   = \frac{G^{(1)}(\mathbf{k}_1,\mathbf{k}_2)}{\sqrt{G^{(1)}(\mathbf{k}_1,\mathbf{k}_1)}  \sqrt{G^{(1)}(\mathbf{k}_2,\mathbf{k}_2)}} \\
& =  \frac{1}{I_\text{tot}} \sum_{i=1}^N \sum_{j=1}^N \langle   e^{- i\mathbf{k}_1  \cdot \mathbf{r}_{i}} e^{-i \phi_i} \, \hat{a}_i^\dagger   \  e^{ i \mathbf{k}_2 \cdot \mathbf{r}_{j}} e^{i \phi_j} \, \hat{a}_j \rangle  \\
& =  \frac{1}{I_\text{tot}}  \sum_{i=1}^N  \langle \hat{n}_i \rangle \, e^{i \mathbf{q}  \cdot \mathbf{r}_{i}} \sim FT\{ S(\mathbf{r}) \} = \tilde{S}(\mathbf{q}) \, ,
\end{aligned}
\label{eq:17}
\end{equation}
where $\mathbf{q} = \mathbf{k}_2 - \mathbf{k}_1$ now refers to the difference between the two outgoing wave vectors $\mathbf{k}_2$ and $\mathbf{k}_1$ (see Figs.~\ref{fig:scattering}c and \ref{fig:scattering}d). If we were to measure $g^{(1)}(\mathbf{k}_1,\mathbf{k}_2)$ it would be possible to extract 3D structural information from an ensemble of incoherently radiating emitters, including the phase \cite{Zarubin1993,Yariv1996a,Yariv1996b,Brady1999,Wolf2009}. Yet, measuring $g^{(1)}(\mathbf{k}_1,\mathbf{k}_2)$ is hard even for macroscopic objects in the visible, and entirely impractical in the x-ray regime. By contrast, considering the spatial (equal-time) second-order intensity correlation function
\begin{equation}
\begin{aligned}
G^{(2)}(\mathbf{k}_1,\mathbf{k}_2) &= \langle E^{(-)}(\mathbf{k}_1) E^{(-)}(\mathbf{k}_2) E^{(+)}(\mathbf{k}_2)  E^{(+)}(\mathbf{k}_1)  \rangle \\
g^{(2)}(\mathbf{k}_1,\mathbf{k}_2) & = \frac{G^{(2)}(\mathbf{k}_1,\mathbf{k}_2)}{G^{(1)}(\mathbf{k}_1,\mathbf{k}_1)  G^{(1)}(\mathbf{k}_2,\mathbf{k}_2)}  \, ,
\end{aligned}
\label{eq:18}
\end{equation}
we obtain for TLS 
\begin{equation}
g^{(2)}_{\text{TLS}}(\mathbf{k}_1,\mathbf{k}_2)  =  1 +  |g^{(1)}(\mathbf{k}_1,\mathbf{k}_2)|^2 \, ,
\label{eq:19}
\end{equation}
known as the \textit{Siegert} relation \cite{Goodman1985}. For SPE we get
\begin{equation}
\begin{aligned}
& g^{(2)}_{\text{SPE}}(\mathbf{k}_1,\mathbf{k}_2)\\
& = \frac{1}{ I_\text{tot}^2} \sum_{i\neq j = 1}^N \Big( e^{- i \mathbf{k}_1  \cdot \mathbf{r}_{i}} e^{- i \mathbf{k}_2  \cdot \mathbf{r}_{j}}  e^{ i \mathbf{k}_2  \cdot \mathbf{r}_{i}} e^{ i \mathbf{k}_1  \cdot \mathbf{r}_{j}}  \\
 & \hspace{18mm} \times  \langle   \hat{a}_i^{\dagger}  \hat{a}_j^{\dagger}  \hat{a}_i \hat{a}_j  \rangle  
+ \langle   \hat{a}_i^{\dagger}  \hat{a}_j^{\dagger}  \hat{a}_j \hat{a}_i  \rangle \Big) \\
& = \frac{1}{ I_\text{tot}^2} \left( \sum_{i\neq j = 1}^N \langle \hat{n}_i \rangle \langle \hat{n}_j \rangle +  \sum_{i\neq j = 1}^N \langle \hat{n}_i \rangle \langle \hat{n}_j \rangle   \ e^{ i \mathbf{q}  \cdot \mathbf{r}_{i}}  e^{- i \mathbf{q}  \cdot \mathbf{r}_{j}} \right) \\
& = 1 - \frac{2}{N} + |g^{(1)}(\mathbf{k}_1,\mathbf{k}_2)|^2 \, ,
\end{aligned}
\label{eq:20}
\end{equation}
where we used $I_\text{tot} \equiv \sum_i \langle \hat{n}_i \rangle$. This shows that measuring the second-order intensity correlation function for TLS or SPE gives indeed access to the 3D Fourier magnitudes $ |\tilde{S}(\mathbf{q})|^2$.  The 3D structure of the arrangement of the emitting species in real space can then be reconstructed by using again well-known phase retrieval algorithms \cite{Fienup1982,Elser2003}.

IDI based on second-order intensity correlation measurements bears several advantages with respect to CDI. The atomic cross sections of incoherent processes like fluorescence emission are generally significantly larger than for coherent ones, producing higher signals compared to CDI. Furthermore, incoherent fluorescence emission displays a uniform angular distribution.  This is unlike conventional crystallography or single-particle CDI where the coherently scattered intensities generally follow a $\mathbf{q}^{-4}$-dependence for small $\mathbf{q}$-values, i.e., at low resolutions \cite{Brown2006}. In addition, considering crystals, the coherently scattered signal is concentrated into Bragg peaks. Both of these features require a high dynamic range of the detectors which can limit the achievable resolution. By contrast, IDI does not require high dynamic range measurements and it is as easy to measure $\mathbf{q}=0$ (auto correlation of each pixel), as it is to measure any other value up to the largest difference of wavevectors captured by the detector.

\begin{figure}[t!]%
\centering
\includegraphics[width=0.80 \linewidth]{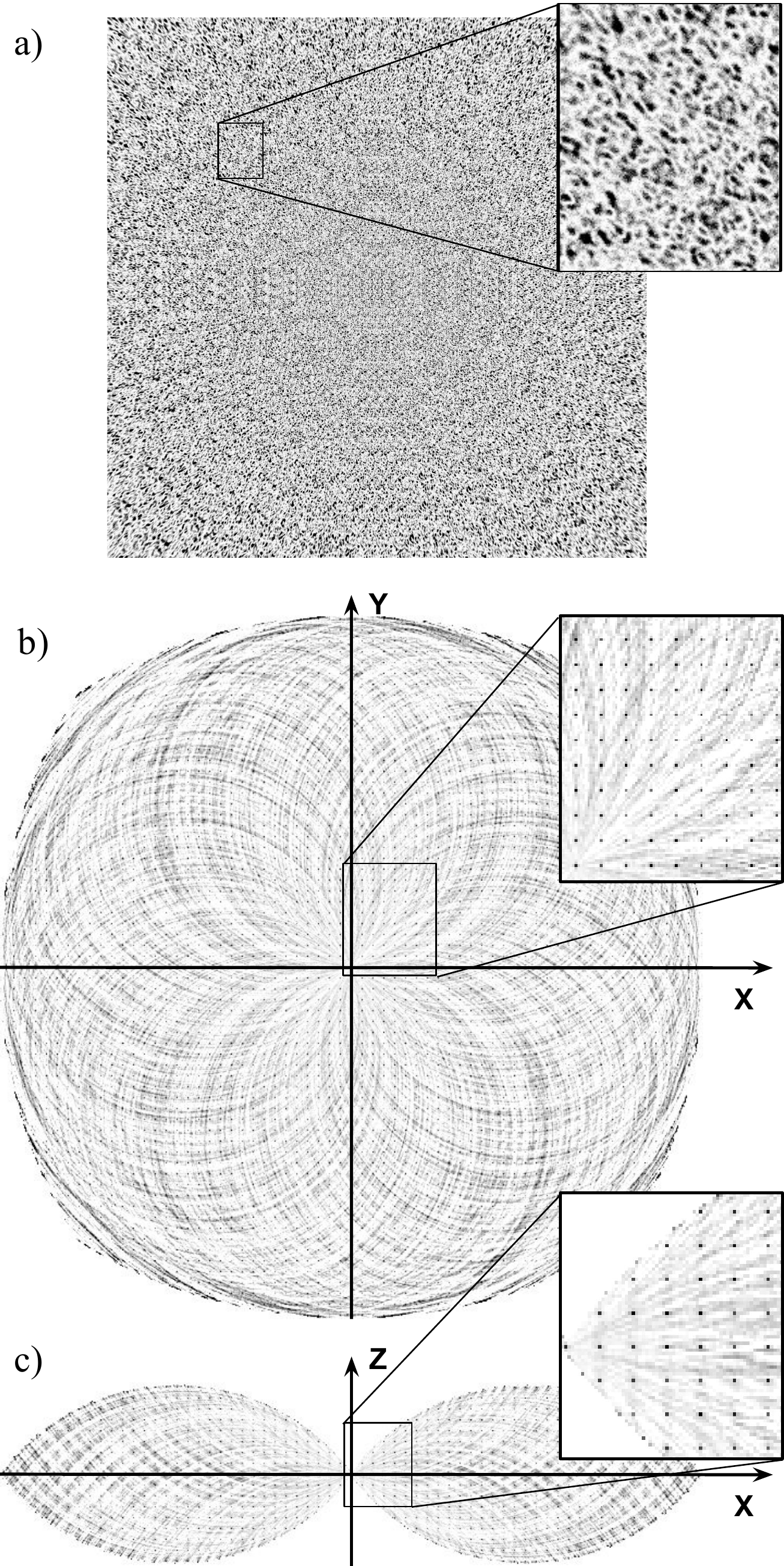}
\caption{Simulation of a mono-atomic cubic crystal with $10\times 10\times 10$ unit cells and a single fluorescing atom per cell showing Bragg peaks in the IDI signal. a) Intensity distribution on the detector resulting from a single $\leq 2.6$\,fs XFEL excitation pulse. The inset shows random speckles with no indication that the object is periodic. b) and c) Orthogonal slices through the $\mathbf{q}$-space intensity autocorrelation. The insets show Bragg peaks corresponding to the lattice constant of the crystal.}
\label{fig:sim}%
\end{figure}

Aside from providing simple access to larger $\mathbf{q}$ vectors IDI doubles the accessible range in Fourier space compared to CDI for the same experimental geometry. This can readily be recognized from the illustrations in Figs.\,\ref{fig:scattering}b and \ref{fig:scattering}d and Figs.\,\ref{fig:sim}b and \ref{fig:sim}c: since all combinations $\mathbf{q}=\mathbf{k}_2-\mathbf{k}_1$ accessible by the pixels of the detector build up the observable region in Fourier space, the largest $\mathbf{q}$ vector reaches out twice as far as in CDI for common detector geometries. Furthermore, IDI leads to volumetric 3D information in Fourier space for a single sample (or detector) orientation, which means that only a few orientations need to be measured to fill the full 3D Fourier space. In contrast, CDI requires fine angular sampling of the probe to build up sufficient completeness. Therefore, amazingly, atomic resolution can be achieved in IDI already with moderate x-ray photon energies and for very few orientations. In addition, since the number of $\mathbf{q}$ vectors obtained from a single frame scales as the square of the number of pixels, binning the resulting $\mathbf{q}$ vectors into a 3D grid results in a large amount of statistics from only a few images, as is well-known from 2D speckle pattern recognition \cite{Katz2014,Schneider2017}.

In order to obtain $g^{(2)}$-signals with high visibility the detection time should be on the order of (or below) the coherence time $\tau_c$ of the photons emanating from the sample. It is the virtue of IDI based on fluorescence emission that the detection time can be intrinsically replaced by the natural time-gating capability of ultra-short x-ray FEL pulses, where the detector needs merely to discriminate between individual pulses and data acquisition needs to keep up with the pulse repetition rate. Typical pulse durations at current x-ray FEL facilities are on the order of 50\,fs in the high bunch-charge mode, whereas low bunch-charge modes already enable pulse durations close to 2\,fs \cite{Ding2015}; this is already shorter than, e.g., the radiative lifetime $\tau_c =  2.6$\,fs of the $K_\alpha$ fluorescence in Fe atoms. Note that good statistics can be achieved rapidly due to the extreme brilliance and high pulse repetition rates of current x-ray FEL facilities. For example, the European XFEL is expected to produce $27,000$ pulses per second, where for IDI a few hundred images may already suffice to obtain high-quality 3D diffraction data, corresponding to sub-second data acquisition times. We point out that the method also works with pulse durations $>\tau_c$, leading, however, to a reduced contrast of the $g^{(2)}$-signal \cite{Yabashi2002}. The latter scales with the ratio of coherence time to time resolution resulting from the integration over independent temporal modes, where correlations of photons of the same mode lead to interferences, whereas correlations of photons of different modes add to the offset. Yet, the reduced visibility can be overcome by averaging over more exposures in order to obtain a sufficient signal to noise ratio (SNR) in the underlying diffraction pattern.

A simulation for a small crystal with $10\times 10\times 10$ unit cells and a single fluorescing atom per cell in a micro-focussed x-ray beam has been performed (see Fig.\,\ref{fig:sim}). A $1745 \times 1745$ pixel detector with 0.11\,mm pixel size was placed 70\,mm from the interaction point. Fig.\,\ref{fig:sim}a displays the single shot intensity distribution from a short $\leq 2.6$\,fs pulse on the detector and Figs.\,\ref{fig:sim}b and \ref{fig:sim}c show orthogonal slices through the reciprocal space intensity autocorrelation evaluated from this single image. In the simulation, the interference from spherical waves with random initial phases from each atom was calculated on each pixel and the intensities were Poisson sampled, where the mean number of photons on the entire detector was $1.7 \times 10^7$ \footnote{This corresponds to a 0.15\,mJ, 7.2\,keV XFEL pulse focused to 1x1\,$\mu$m$^2$ incident on a 1\,$\mu$m$^3$ ferredoxin protein crystal (PDB: 1FDN)}. Even though the intensity distribution on the detector does not seem to contain any information, the intensity autocorrelation depicted in Figs.\,\ref{fig:sim}b and \ref{fig:sim}c shows significant Bragg peaks related to the crystalline order of the sample just from the single exposure. For lower intensities Bragg peaks may not rise above noise from the evaluation of a single shot, but averaging over many exposures will rapidly increase the SNR.

Finally, we note that  fluorescence-based IDI enables element-specific imaging where by use of appropriate energy filters different species in the same or different molecules can be selectively resolved. As such, IDI can be combined with CDI, where IDI is recorded at a scattering angle of $90^\circ$ (where stray light and coherently scattered radiation is highly suppressed) and CDI is recorded simultaneously with a second detector in the forward direction. For the reconstruction of the sample the IDI signal can provide particular atom positions with very high resolution, which then can be used to phase the CDI Fourier amplitudes of large macromolecular proteins in a manner similar to anomalous dispersion techniques \cite{Hendrickson1981,Hendrickson1991}.

In conclusion, we presented a novel diffractive imaging technique, incoherent diffraction imaging (IDI), which---based on the measurement of intensity correlations in the far-field---allows to extract 3D structural information from incoherently emitting objects with atomic resolution. Like CDI, IDI gives access to the modulus of the 3D scattering amplitudes, yet with twice the resolution compared to CDI for common detector geometries and additional volumetric information in Fourier space for a single sample orientation. The requirements for the implementation---high brilliance, ultra-short excitations and high repetition rates with detectors keeping up---are ideally met by current FEL facilities, making IDI a timely and cutting-edge technique with the potential to substantially improve x-ray structure determination. 
A prospective important application would be to image metal-bearing clusters in metalloproteins where the clusters mediate reactivities and functions that are of fundamental importance for the biosphere on Earth. Examples are Iron-sulphur clusters \cite{Lindahl1990}, e.g., in nitrogenases that are responsible for nitrogen fixation \cite{Rees2014}, or the Mn$_4$Ca cluster in Photosystem II that catalyzes the water oxidation reaction in photosynthesis \cite{Yachandra2014}. Common to all these reactions is that they are accompanied by subtle, yet unresolved structural changes of these clusters which could be revealed with ultrahigh resolution via IDI. The method may also provide a new route to achieve single-molecule diffractive imaging \cite{Neutze2000}, which currently suffers from insufficient signal, excessive background, and difficulty in delivering molecules ``container free'' to the X-ray beam \cite{Aquila2015}. IDI targeted towards sulphur or phosphorous atoms, on the other hand, provides significantly more signal for a given incident intensity, and zero background even when delivering samples in a water micro-jet \cite{Lee2016} aimed at the focused X-ray beam.

\begin{acknowledgments}

K.A, H.N.C and R.R. acknowledge the support of the Helmholtz Association through project-oriented funds. A.C. and J.v.Z. gratefully acknowledge funding by the Erlangen Graduate School in Advanced Optical Technologies (SAOT) by the German Research Foundation (DFG) in the framework of the German excellence initiative.

\end{acknowledgments}

\end{document}